\documentclass[showpacs,aps,pra,superscriptaddress]{revtex4}

\usepackage{amsmath}
\usepackage{amssymb,dsfont}
\usepackage[dvips]{graphicx}

\begin{document}

\newcommand{\imu}{\textrm{i}}

 \title{Adiabatic entanglement in two-atom cavity QED}

\author{C.~Lazarou}
\author{B.M.~Garraway}
 \affiliation{Department of Physics and Astronomy,
 University of Sussex, Falmer, Brighton, BN1 9QH, United Kingdom}


\date{\today}

\begin{abstract}
We analyse the problem of a single mode field
interacting with a pair of two level atoms. The atoms
enter and exit the cavity at different times. Instead
of using constant coupling, we use time dependent
couplings which represent the spatial dependence of
the mode. Although the system evolution is adiabatic
for most of the time, a previously unstudied energy crossing plays a key
role in the system dynamics when the atoms have a time delay. We show that conditional
atom-cavity entanglement can be generated, while for
large photon numbers the entangled system has a behaviour
which can be mapped onto the single atom Jaynes-Cummings
model. Exploring the main features of this system we
propose simple and fairly robust methods for entangling atoms independently of the
  cavity, for 
quantum state mapping, and for implementing SWAP and C-NOT gates with
  atomic qubits.
\end{abstract}


\maketitle

\section{Introduction} \label{sec:Intro}
In recent years many authors have proposed different schemes, based
on cavity QED systems, for entangling atoms and implementing quantum
logic gates. Some of these proposals use single resonant
interactions \cite{Zheng2005} or strongly detuned cavities
\cite{Zheng2000,Jane2002,You2003b} for entangling atoms or carrying
out logic gates. More elaborate schemes use decoherence-free spaces
and continuous monitoring of the cavity decay for generating
entangled states \cite{Plenio1999,Beige2000a,Beige2000b}. For these
cases a no-photon emission is associated with the generation of a
maximally entangled state. In the optical cavity regime \cite{Miller2005}, 
photon polarisation measurements are used for entangling atoms \cite{Duan2003} or
implementing quantum logic gates \cite{Duan2004,Duan2005}.
With these theoretical proposals the desired outcome is achieved by single-photon pulse
scattering by an atom trapped inside an optical cavity \cite{Duan2003A}.

As well as these theoretical proposals a number of experiments entangling Rydberg
  atoms in microcavities have been carried out
  \cite{Raimond2001,Hagley1997,Osnaghi2001,Rauschenbeutel2000,Rauschenbeutel1999,Nogues1999}.
  EPR pairs \cite{Raimond2001,Hagley1997,Osnaghi2001} and entangled states of
  three quantum systems \cite{Raimond2001,Rauschenbeutel2000} were generated,
  a controlled phase
  gate and a C-NOT gate between a photon and an atom
  \cite{Raimond2001,Rauschenbeutel1999,Nogues1999} were implemented, a single photon
  non-demolition detection \cite{Rauschenbeutel2000} was performed, and entanglement between an
  atom and a mesoscopic cavity field and measurement of decoherence effects
  \cite{Raimond2001} were carried out. Most of these experiments are based on the single atom
  Jaynes-Cummings \cite{Scully} model where the atom and the cavity field, with zero or one
  photon, represent the qubits. The only exception is an experiment based on
  atomic collisions inside a strongly detuned cavity
  \cite{Zheng2000,Osnaghi2001} for which the two atoms which collide inside
  the cavity are also the qubits. This experiment, in common with the proposal we discuss here, is
  also relatively simple in that it does not require the use of
  Ramsey zones. So in contrast to many of these proposals,
  in this paper we aim to obtain entangled atoms where the entanglement is not
  ultimately complicated by additional cavity entanglement.

Among the theoretical proposals we also find the adiabatic, sequential
passage of pairs of atoms through cavities \cite{Marr2003,Yong2007}.
The scheme discussed by Marr et.\ al.\ \cite{Marr2003} exploits features
similar to those appearing in STIRAP
\cite{Bergmann1995,Bergmann1998}. The system adiabatically follows a
dark state and a maximally entangled state between two atoms can be
generated. In addition to this, Yong, Bruder and Sun \cite{Yong2007}
proposed a scheme for generating entangled atomic states by
sequential passage of atoms through a strongly detuned cavity. Using
a time dependent Fr\"ohlich transformation, they derived an
effective Hamiltonian and consider the preparation of entangled
states with respect to the atomic velocities and the initial
displacement between the atoms. We will see that the system we study here neither
relies on the detection of lost photons, as Marr et.~al. do \cite{Marr2003},
nor requires accurate control of the delay between the atoms, as in the case
of Yong
et.~al. \cite{Yong2007}.

We consider here as a key component of our scheme, the adiabatic
limit for a system of two atoms
resonantly coupled to a single mode field. Instead of the usual
assumption of a constant coupling between the atoms and the field,
we will utilise sequential time dependent couplings, which reflect
the possibility that the two atoms enter the cavity mode at
different times. From the analysis that follows, we show that the
system is characterised by an energy crossing. This, and the
symmetric structure of the adiabatic spectrum, make the system
\emph{fairly} robust; in an experiment one has to control a
single mixing angle. Furthermore we show that a conditional
atom-cavity entanglement can be generated and demonstrate a
connection to the Jaynes-Cummings model \cite{Scully} which occurs
when the cavity is highly excited. We also deduce the necessary
conditions for using the adiabatic approximation \cite{Messiah}, and
discuss the feasibility of a potential experiment. Finally using our
results we describe methods for entangling atoms and for mapping
quantum states, and we propose two setups for implementing a SWAP and a
control-NOT gate (C-NOT). The proposed applications are fairly
robust and simple to implement, whereas the operations could be
relatively fast preventing potential failures due to decoherence
effects.

In contrast to previously proposed
  \cite{Zheng2005,Zheng2000,Jane2002,You2003b,Plenio1999,Beige2000a,Beige2000b,Duan2003,Duan2004,Duan2005,Duan2003A}
  or demonstrated methods
  \cite{Raimond2001,Hagley1997,Osnaghi2001,Rauschenbeutel2000,Rauschenbeutel1999,Nogues1999}
  for entangling quantum systems or implementing quantum logic gates, here we
  consider a resonant two-atom, time-dependent Tavis-Cummings model, which
  differs from previous proposals either because they use a single
  atom interacting with a cavity, or because the interactions between the atom
  or atoms and the cavity are assumed to be constant in time. In addition to this, the main
  difference is that the resulting dynamics are different from what one
  expects and this is because of the existence of the energy crossing which is
  something new, and to our knowledge has never been reported before. 

The paper is organised as follows. In section \ref{sec:Model} we discuss our model and introduce the corresponding
interaction Hamiltonian. After deriving the adiabatic states, we move to section \ref{sec:Results} were we present
and analyse our results. In section \ref{sec:4} we discuss potential
applications for quantum information processing and quantum state mapping. We conclude by summarising our
results in section \ref{sec:Conl}.

\section{The atom-cavity model}\label{sec:Model}
\subsection{The Hamiltonian} \label{sec:Hamiltonian}
\label{sec:sect2}

Our system consists of a pair of two-level atoms which resonantly
interact with a single mode cavity field via time dependent
couplings. In the interaction picture, and within the rotating wave
approximation, the Hamiltonian reads $(\hbar=1)$
\begin{equation} \label{eq:1}
H_{I}(t)=\sum_{j=1}^{2}\eta_{j}(t)\big(a^{\dagger}\sigma^{j}_{-}+a\sigma^{j}_{+}\big)
.
\end{equation}
The operators $a^{\dagger}$ and $a$ are the Bosonic
creation-annihilation operators  for the cavity mode, and
$\sigma^{j}_{-}$ and $\sigma^{j}_{+}$ are the lowering-raising
operators for atom $j$.

Both atoms are considered to traverse the cavity with the same
speed $v$, following the same trajectory $x(t)$, but atom 2 is
delayed by $2\Delta t$ with respect to atom 1. The field
spatial profile along the trajectory $x(t)$ is assumed to be Gaussian
\begin{equation} \label{eq:2}
E(x)=E_{0}\exp\big(-\frac{x^{2}}{4x_{0}^2}\big).
\end{equation}
The use of these assumptions results in Gaussian coupling pulses $\eta_{j}(\tau)$:
\begin{equation} \label{eq:3}
\eta_{\scriptscriptstyle
1}(\tau)=ge^{-(\tau+\delta)^{2}},\qquad\eta_{\scriptscriptstyle
2}(\tau)=ge^{-(\tau-\delta)^{2}}.
\end{equation}
Here
the dimensionless time $\tau$ and the parameter $\delta$ are defined
in terms of the time width $\sigma=x_{0}/v$
\begin{equation} \label{eq:4}
\tau=\frac{t}{2\sigma},\qquad\delta=\frac{\Delta t}{2\sigma}.
\end{equation}
The width $\sigma$ could be seen as an average atom-cavity
interaction time and later we will find that it can play the role of the interaction time in the
Jaynes Cummings model \cite{Scully}.

When deriving the Hamiltonian (\ref{eq:1}), we assumed that the atoms
are moving sufficiently fast that the kinetic energy for each
atom is greater than its coupling strength $\eta_{j}(\tau)$. For this
limit the atomic momentum operator $\hat{p}$ and the displacement operator
$\hat{x}$ can be replaced by their classical counterparts $mv$ and $vt$
respectively. This requirement is easily satisfied with atomic beams, and we
will also not expect the results of the variations to affect the form of the
coupling shapes $\eta_{j}(\tau)$. However, we not that in the opposite limit, when the atoms 
are slow, the atoms
could be reflected without traversing the cavity \cite{Meyer1997}. 

In what follows we consider a parameter regime in which the
adiabatic theorem holds \cite{Messiah}. Then if the
Hamiltonian of the system is a smooth and slowly varying function of
time, and the initial state of the system is one of the adiabatic
states $\left\vert\Psi_{j}(t_{i})\right\rangle$ of $H_{I}(t)$, where
\begin{equation} \label{eq:5}
H_{I}(t)\left\vert\Psi_{j}(t)\right\rangle=E_{j}(t)\left\vert\Psi_{j}(t)\right\rangle,
\end{equation}
then at subsequent times $t_{f}$, the system will be found in the
adiabatic state $\left\vert\Psi_{j}(t_{f})\right\rangle$
\begin{equation} \label{eq:6}
\left\vert\Psi(t_{f})\right\rangle=e^{-\imu\phi_{j}(t_{f})}\left\vert\Psi_{j}(t_{f})\right\rangle.
\end{equation}
Eq. (\ref{eq:6}) is usually referred as the adiabatic
approximation \cite{Messiah}, where the dynamical phase acquired
during the system evolution is
\begin{equation} \label{eq:7}
\phi_{j}(t)=\int_{t_{i}}^{t}dt' E_{j}(t').
\end{equation}
The conditions for applying this approximation will be discussed in detail later.
We should also mention that the adiabatic theorem assumes that the
Hamiltonian has a discrete spectrum, and for degeneracies the
adiabatic approximation could fail.

\subsection{Adiabatic states} \label{sec:statenergie}
The Hamiltonian $H_{I}(\tau)$, Eq. (\ref{eq:1}), couples only
bare states with the same number of total excitations. Thus our analysis
is restricted to a subspace spanned by four states of the initial vector
space, i.e.
\begin{equation} \label{eq:8}
\vert n,e_{1}e_{2}\rangle,\quad\vert n+1,g_{1}e_{2}\rangle,\quad\vert
n+1,e_{1}g_{1}\rangle,\quad\vert n+2,g_{1}g_{2}\rangle .
\end{equation}
The excited state of atom $j$ is denoted $\vert e_{j}\rangle$,
$\vert g_{j}\rangle$ is the ground state, and $\vert
n\rangle$ refers to the field state with $n$ excitations.

The adiabatic energies are the roots of the characteristic polynomial
$P(E)$
\begin{equation} \label{eq:9}
P(E)=det\vert \mathds{H}_{I}(\tau)-E\vert,
\end{equation}
where $\mathds{H}_{I}$ is the matrix representation of the
Hamiltonian in the subspace (\ref{eq:8}). From Eqs.
(\ref{eq:1}) and (\ref{eq:8}) we have
\begin{equation} \label{eq:10}
P(E)=E^{4}-E^{2}(3+2n)(\eta_{\scriptscriptstyle
1}^{2}+\eta_{\scriptscriptstyle
2}^{2})+(1+n)(2+n)(\eta_{\scriptscriptstyle
1}^{2}-\eta_{\scriptscriptstyle 2}^{2})^{2}.
\end{equation}
Because $P(E)$ includes only even powers of $E$, the derivation of
the corresponding roots is simple. The resulting adiabatic energies
are
\begin{subequations} \label{eq:11}
\begin{align}
E_{1,2}(\tau)&=\mp E_{-}(\tau),\qquad E_{3,4}=\mp E_{+}(\tau),
\label{eq:11a} \\ \nonumber \\
E_{\pm}(\tau)&=\sqrt{\frac{(3+2n)(\eta_{\scriptscriptstyle
1}^{2}(\tau)+\eta_{\scriptscriptstyle 2}^{2}(\tau))\pm
F_{n}(\tau)}{2}}, \label{eq:11b}
\end{align}
\end{subequations}
where the function $F_{n}(\tau)$ is
\begin{equation} \label{eq:12}
F_{n}(\tau)=\sqrt{\big(\eta_{\scriptscriptstyle
1}^{2}(\tau)+\eta_{\scriptscriptstyle
2}^{2}(\tau)\big)^{2}+16(n+1)(n+2)\eta_{\scriptscriptstyle
1}^{2}(\tau)\eta_{\scriptscriptstyle 2}^{2}(\tau)}.
\end{equation}

With the energies in hand, Eq. (\ref{eq:5}) can be solved to
give the adiabatic states,
\begin{subequations} \label{eq:13}
\begin{align}
\nonumber \left\vert\Psi_{1,2}(\tau)\right\rangle =&
A_{-}(\tau)\vert n,e_{1}e_{2}\rangle+D_{-}(\tau)\vert
n+2,g_{1}g_{2}\rangle \\ &\pm \big(B_{-}(\tau)\vert
n+1,g_{1}e_{2}\rangle-C_{-}(\tau)\vert n+1,e_{1}g_{2}\rangle\big),
\label{eq:13a} \\ \nonumber
\left\vert\Psi_{3,4}(\tau)\right\rangle =& A_{+}(\tau)\vert
n,e_{1}e_{2}\rangle+D_{+}(\tau)\vert n+2,g_{1}g_{2}\rangle \\ &\pm \big(B_{+}(\tau)\vert
n+1,g_{1}e_{2}\rangle-C_{+}(\tau)\vert n+1,e_{1}g_{2}\rangle\big).
\label{eq:13b}
\end{align}
\end{subequations}
The upper sign in Eqs. (\ref{eq:13}) is for the odd numbered
states and the lower one is for the even ones.
The coefficients $A_{\pm}(\tau), B_{\pm}(\tau), C_{\pm}(\tau)$ and
$D_{\pm}(\tau)$ are given in the appendix and Ref.\ \cite{Mahmood1987}.
For the purposes of our analysis it is sufficient to derive the limits of the adiabatic states for $\tau\rightarrow\pm\infty$ and $\tau=0$.

\begin{figure}[!t]
    \centerline{\includegraphics{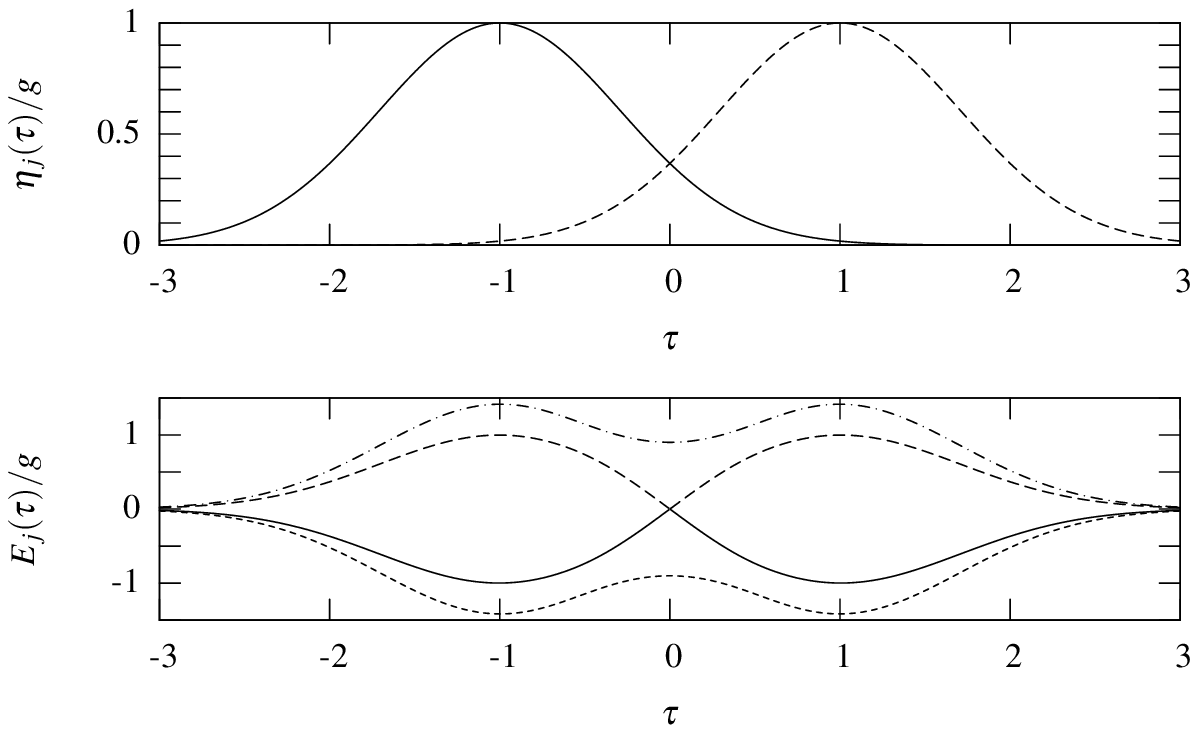}}
 \caption{Top: The coupling functions $\eta_{\scriptscriptstyle 1}(\tau)$ (solid) and $\eta_{\scriptscriptstyle 2}(\tau)$ (dashed) with respect to $\tau$. Bottom: The adiabatic energies $E_{1}(\tau)$ (solid), $E_{2}(\tau)$ (long dashed), $E_{3}(\tau)$ (dashed) and $E_{4}(\tau)$ (chain) with respect to $\tau$. The parameters are $\delta=1.0$ and $n=0$.
  \label{fig:1}}
\end{figure}

For the coupling functions (\ref{eq:3}), and for $\tau=\pm\infty$,
we assume that
\begin{displaymath}
\lim_{\tau\rightarrow\infty}\left(\frac{\eta_{\scriptscriptstyle
1}}{\eta_{\scriptscriptstyle
2}}\right)=0,\qquad\lim_{\tau\rightarrow-\infty}\left(\frac{\eta_{\scriptscriptstyle
2}}{\eta_{\scriptscriptstyle 1}}\right)=0.
\end{displaymath}
Once these relations are used with Eqs. (\ref{eq:13a}) and
(\ref{eq:13b}), the adiabatic states for both limits are derived
\begin{subequations} \label{eq:14}
\begin{align}
\left\vert\Psi_{1,2}(\tau)\right\rangle&=-\frac{1}{\sqrt{2}}\left\{
                                                       \begin{array}{ll}
                                                         \vert n,e_{1}e_{2}\rangle\mp\vert n+1,g_{1}e_{2}\rangle& \textrm{for}\quad \tau\rightarrow-\infty
                                                           \\
                                                         & \\
                                                         \vert n,e_{1}e_{2}\rangle\mp\vert n+1,e_{1}g_{2}\rangle& \textrm{for}\quad\tau\rightarrow\infty
                                                       \end{array}
                                                     \right.
\label{eq:14a} \\ \nonumber \\
\left\vert\Psi_{3,4}(\tau)\right\rangle&=-\frac{1}{\sqrt{2}}\left\{
                                                       \begin{array}{ll}
                                                         \vert n+2,g_{1}g_{2}\rangle\mp\vert n+1,e_{1}g_{2}\rangle& \textrm{for}\quad \tau\rightarrow-\infty
                                                           \\
                                                         & \\
                                                         \vert n+2,g_{1}g_{2}\rangle\mp\vert n+1,g_{1}e_{2}\rangle& \textrm{for}\quad\tau\rightarrow\infty
                                                       \end{array}
                                                     \right.
\label{eq:14b}
\end{align}
\end{subequations}

From Eqs. (\ref{eq:14a}) and (\ref{eq:14b}) we see that the
adiabatic states, do not match individual bare states in both limits. This is
well justified if we take into account the fact that the bare states
are degenerate. Because of this even a small interaction is enough
to lift the degeneracy and mix the bare states at the early
stages of the system evolution.

For $\tau=0$, we have that $E_{1}(0)=0=E_{2}(0)$. This can be seen
either by putting $\eta_{\scriptscriptstyle
1}=\eta_{\scriptscriptstyle 2}$ in (\ref{eq:11}), or from figure
\ref{fig:1}. The corresponding degenerate states are also
discontinuous functions of $\tau$
\begin{equation} \label{eq:15}
\left\vert\Psi_{1,2}(\tau\rightarrow0^{\pm})\right\rangle=-\beta_{n}\vert
n,e_{1}e_{2}\rangle+\alpha_{n}\vert
n+2,g_{1}g_{2}\rangle\pm\frac{\textrm{sign}(\tau)}{\sqrt{2}}\vert-\rangle.
\end{equation}
The coefficients $\alpha_{n}$, $\beta_{n}$ and the state
$\vert-\rangle$ are
\begin{subequations} \label{eq:16}
\begin{align}
\alpha_{n}=\sqrt{\frac{1+n}{6+4n}},\qquad\beta_{n}=\sqrt{\frac{n+2}{6+4n}},
\label{eq:16a}
\\ \vert\pm\rangle=\frac{1}{\sqrt{2}}\vert
n+1\rangle\big(\vert g_{1}e_{2}\rangle\pm\vert
e_{1}g_{2}\rangle\big). \label{eq:16b}
\end{align}
\end{subequations}
The remaining adiabatic
states are continuous with respect to $\tau$
\begin{equation} \label{eq:17}
\left\vert\Psi_{3,4}(0)\right\rangle=\alpha_{n}\vert n,e_{1}e_{2}\rangle+\beta_{n}\vert n+2,g_{1}g_{2}\rangle\mp\frac{1}{\sqrt{2}}\vert+\rangle.
\end{equation}

Thus we see that the Hamiltonian $H_{I}(\tau)$ has a temporal degeneracy. As already mentioned, when encountering such
degeneracies the adiabatic approximation is expected to fail. If the degenerate states are coupled, then
the system is likely to be found in a superposition of these states for $\tau>0$, even if the coupling is
relatively weak. A solution to this problem could be given by using a perturbation series \cite{Hagedorn1989}. Then, the
resulting state vector will be expressed in terms of both adiabatic states, through the various terms appearing in the
perturbation expansion. Here, instead of using a perturbation series, a
simpler approach is considered based on an adiabatic elimination of two of
the states.
\section{System dynamics} \label{sec:Results}

\subsection{Energy crossing} \label{sec:3_1}

In order to predict the system evolution for $\tau\sim0$ we expand the Hamiltonian near the temporal degeneracy.
From this expansion an effective Hamiltonian is obtained.
The first step of this approach, is to change the current basis into that of the
adiabatic states for $\tau=0^{-}$,
\begin{eqnarray}
\nonumber
H_{I}(\tau)&=&\sum_{j=1,2}(-1)^{j}\omega_{1}\Omega_{-}(\tau)\vert\psi_{j}\rangle\langle\psi_{j}\vert+
\sum_{j=3,4}(-1)^{j}\omega_{2}\Omega_{+}(\tau)\vert\psi_{j}\rangle\langle\psi_{j}\vert
\\ & &+\frac{\Omega_{-}(\tau)}{4\omega_{2}}\big(\vert\psi_{2}\rangle-\vert\psi_{1}\rangle\big)\big(\langle\psi_{3}\vert+\langle\psi_{4}\vert\big)+\textrm{h.c.}.
\label{eq:18}
\end{eqnarray}
The couplings $\Omega_{\pm}$ and the parameters $\omega_{1,2}$ are
\begin{subequations} \label{eq:19}
\begin{align}
&\Omega_{\pm}(\tau)=\eta_{\scriptscriptstyle
1}(\tau)\pm\eta_{\scriptscriptstyle 2}(\tau),
\label{eq:19a}
\\ \nonumber \\
&\omega_{1}=\frac{1}{2}\sqrt{\frac{(n+1)(n+2)}{6+4n}},\qquad\omega_{2}=\frac{1}{2}\sqrt{6+4n},
\label{eq:19b}
\end{align}
\end{subequations}
and $\vert\psi_{j}\rangle=\left\vert\Psi_{j}(0^{-}\right\rangle$.
Then the Schr\"odinger equation in the new basis gives
\begin{subequations} \label{eq:20}
\begin{align}
\imu
\dot{c}_{1}(\tau)=&-\Omega_{-}(\tau)\left(\omega_{1}c_{1}(\tau)+\frac{c_{3}(\tau)+c_{4}(\tau)}{4\omega_{2}}\right),
\label{eq:20a}
\\ \nonumber \\
\imu
\dot{c}_{2}(\tau)=&\Omega_{-}(\tau)\left(\omega_{1}c_{2}(\tau)+\frac{c_{3}(\tau)+c_{4}(\tau)}{4\omega_{2}}\right),
\label{eq:20b}
\\ \nonumber \\
\imu
\dot{c}_{3}(\tau)=&-\omega_{2}\Omega_{+}(\tau)c_{3}(\tau)+\frac{\Omega_{-}(\tau)}{4\omega_{2}}\big(c_{2}(\tau)-c_{1}(\tau)\big),
\label{eq:eq20c}
\\ \nonumber \\
\imu\dot{c}_{4}(\tau)=&\omega_{2}\Omega_{+}(\tau)c_{4}(\tau)+\frac{\Omega_{-}(\tau)}{4\omega_{2}}\big(c_{2}(\tau)-c_{1}(\tau)\big).
\label{eq:eq20d}
\end{align}
\end{subequations}
Up to this point everything is exact. We now note that
$\vert\Omega_{+}(\tau\sim0)\vert\gg\vert\Omega_{-}(\tau\sim0)\vert$. This allows us to
adiabatically eliminate $\vert\psi_{3}\rangle$ and
$\vert\psi_{4}\rangle$, and derive an effective
Hamiltonian. Setting $\dot{c_{3}}=0=\dot{c_{4}}$, and solving for
$c_{3,4}$ in terms of $c_{1}$ and $c_{2}$, results in two
differential equations
\begin{equation} \label{eq:21}
\imu\dot{c}_{1}(\tau)=-\omega_{1}\Omega_{-}(\tau)c_{1}(\tau),\qquad\imu\dot{c}_{2}(\tau)=\omega_{1}\Omega_{-}(\tau)c_{2}(\tau).
\end{equation}
This equation corresponds to an
effective Hamiltonian
\begin{equation} \label{eq:22}
H_{eff}(\tau)=-4\omega_{1}\Omega_{-}(\tau)\hat{\sigma}_{z}\quad\textrm{for}\quad
\tau\sim0,
\end{equation}
where $\hat{\sigma}_{z}$ is the Pauli matrix \cite{Messiah}.

The condition $\vert\Omega_{+}(\tau)\vert\gg\vert\Omega_{-}(\tau)\vert$ holds
for $\vert\tau\vert\ll0.25\vert\delta\vert$, with $\vert\delta\vert\sim1$. For
$\tau\ll-0.25\vert\delta\vert$ we expect the adiabatic approximation
to hold since the energy splittings between the adiabatic states are
relatively large, see section \ref{sec:3_3}. Assuming that the initial state of the system is
$\left\vert\Psi_{1}(-\infty)\right\rangle$, then for
$\tau<-0.25\vert\delta\vert$ the system state reads
\begin{equation} \label{eq:23}
\left\vert\Psi(\tau)\right\rangle=\exp\left(-\imu\int_{-\infty}^{\tau}d\tau'E_{1}(\tau')\right)\left\vert\Psi_{1}(\tau)\right\rangle.
\end{equation}
From Eq. (\ref{eq:22}) we see that $\left\vert\Psi_{1}(0^{-})\right\rangle$ does not couple to
$\left\vert\Psi_{2}(0^{-})\right\rangle$. Thus for $\tau\sim0$ the system state will be
\begin{equation} \label{eq:24}
\left\vert\Psi(0^{-})\right\rangle=\exp\left(-\imu\int_{-\infty}^{0}d\tau'E_{1}(\tau')\right)\left\vert\Psi_{1}(0^{-})\right\rangle\quad\tau\sim0.
\end{equation}
This is true even for $\tau=0^{+}$, but since $\left\vert\Psi_{1}(0^{-})\right\rangle=\left\vert\Psi_{2}(0^{+})\right\rangle$,
Eq. (\ref{eq:15}), the state vector for $\tau>0$ will be
\begin{equation} \label{eq:25}
\left\vert\Psi(\tau)\right\rangle=\exp\left(-\imu\int_{-\infty}^{0}d\tau'E_{1}(\tau')-\imu\int_{0}^{\tau}d\tau'E_{2}(\tau')\right)\left\vert\Psi_{2}(\tau)\right\rangle\quad\tau>0.
\end{equation}

Thus for $\tau=0$ the system undergoes an energy crossing. The consequence of this is that if the system is initially prepared in
$\left\vert\Psi_{1}(-\infty)\right\rangle$, it will end in state $\left\vert\Psi_{2}(\infty)\right\rangle$. Furthermore, since
$E_{1}(\tau)=-E_{2}(-\tau)$ we can prove that
\begin{equation} \label{eq:26}
\int_{-\infty}^{0}d\tau'E_{1}(\tau')+\int_{0}^{\infty}d\tau'E_{2}(\tau')=0
\end{equation}
which means that the net dynamical phase for this pair of adiabatic states will be zero, i.e.
\begin{equation} \label{eq:27}
\left\vert\Psi_{1}(-\infty)\right\rangle\rightarrow\left\vert\Psi_{2}(\infty)\right\rangle,\quad
\left\vert\Psi_{2}(-\infty)\right\rangle\rightarrow\left\vert\Psi_{1}(\infty)\right\rangle.
\end{equation}
In view of these results we can redefine the first two adiabatic states in order to properly incorporate the
energy crossing so that
\begin{subequations} \label{eq:28}
\begin{align}
\left\vert\Psi'_{1}(\tau)\right\rangle&=\theta(-\tau)\left\vert\Psi_{1}(\tau)\right\rangle+\theta(\tau)\left\vert\Psi_{2}(\tau)\right\rangle
\label{eq:28a} \\
\left\vert\Psi'_{2}(\tau)\right\rangle&=\theta(-\tau)\left\vert\Psi_{2}(\tau)\right\rangle+\theta(\tau)\left\vert\Psi_{1}(\tau)\right\rangle
\label{eq:28b}
\end{align}
\end{subequations}
where $\theta(\tau)$ is the unit step function
\begin{equation} \label{eq:29}
\theta(\tau)=\left\{\begin{array}{lll}
1 & \textrm{for}\quad \tau>0 \\
0 & \textrm{for}\quad \tau<0 \\
\frac{1}{2} & \textrm{for}\quad \tau=0.
\end{array}\right.
\end{equation}
The corresponding energies are
\begin{equation} \label{eq:30}
E'_{1}(\tau)=-\textrm{sign}(\tau)E_{-}(\tau),\quad E'_{2}(\tau)=\textrm{sign}(\tau)E_{-}(\tau).
\end{equation}

\begin{figure}[!t]
\begin{center}
\includegraphics{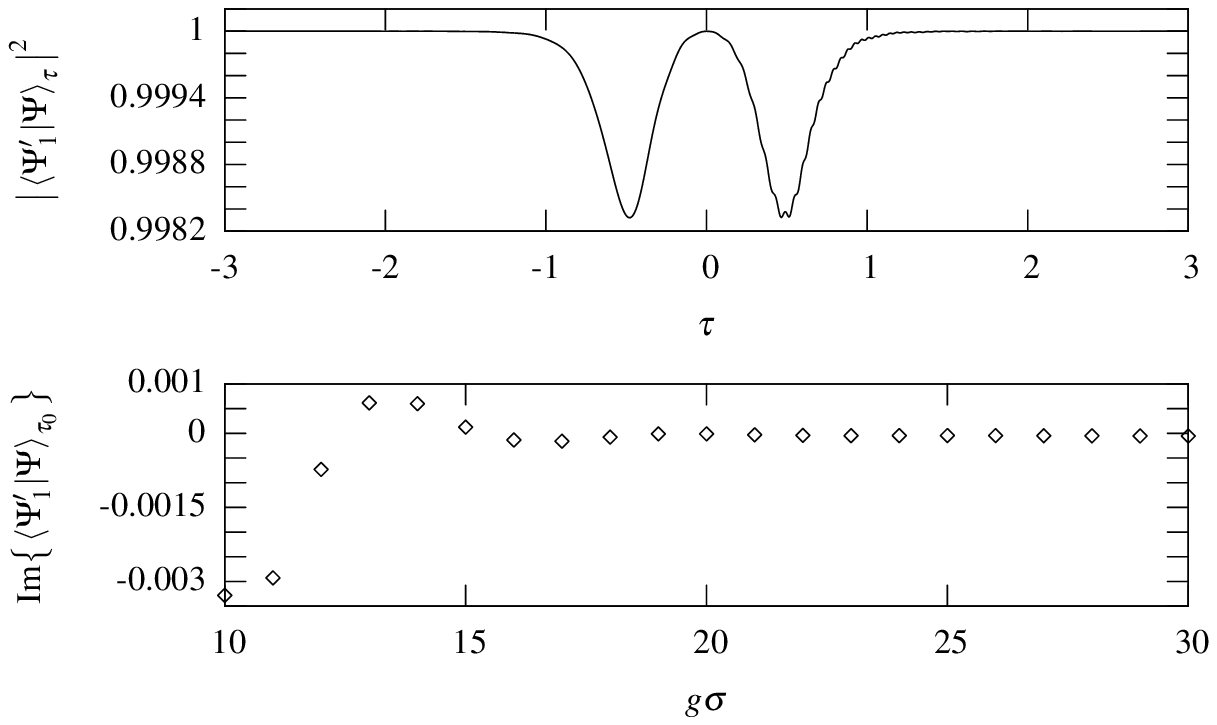}
\end{center}
\caption{Top: The projection
$\left\vert\left\langle\Psi'_{1}(\tau)\vert\Psi(\tau)\right\rangle\right\vert^{2}$
with respect to time. The parameters are $n=0$, $\delta=1.0$ and
$g\sigma=30$. The initial state was
$\left\vert\Psi'_{1}(-\tau_{\scriptscriptstyle 0})\right\rangle$,
and $\left\vert \Psi(\tau)\right\rangle$ is the solution for the
Schr\"odinger equation. Note the narrow range of the vertical axis.
Bottom: The imaginary part of
$\left\langle\Psi'_{1}(\tau_{\scriptscriptstyle 0})
\vert\Psi(\tau_{\scriptscriptstyle 0})\right\rangle$ with respect to
$g\sigma$. The parameters are $n=0$ and $\delta=1.0$.}\label{fig:2}
\end{figure}

To test the validity of Eqs. (\ref{eq:27}), (\ref{eq:28a}) and
(\ref{eq:28b}) we numerically solved the Schr\"odinger equation
within a symmetrical interval $-\tau_{\scriptscriptstyle
0}\leq\tau\leq\tau_{\scriptscriptstyle 0}$. In figure \ref{fig:2}
the projection of Eq. (\ref{eq:28a}) onto the state vector is
plotted with respect to time. From this we see that the system
adiabatically follows state
$\left\vert\Psi'_{1}(\tau)\right\rangle$, with minor non-adiabatic
effects. Despite this, the system ends in state
$\left\vert\Psi_{2}(\tau_{\scriptscriptstyle 0})\right\rangle$ in
accordance with Eq. (\ref{eq:27}). In the lower graph the
imaginary part of the projection
$\left\langle\Psi'_{1}(\tau_{\scriptscriptstyle
0})\vert\Psi(\tau_{\scriptscriptstyle 0})\right\rangle$ is plotted
with respect to the parameter $g\sigma$. This is very small, in
agreement with the expected zero value for the dynamical phase. The
small deviations from zero can be related to minor non-adiabatic
effects during the evolution. The choice made for the system
parameters will be discussed later.

\subsection{Input-output in terms of the bare states} \label{sec:3_2}
For practical purposes, such as quantum information processing, one will encounter the bare states as
input-output states instead of the adiabatic ones. Thus the results of section \ref{sec:3_1} must be expressed in
terms of the bare states. Using Eqs. (\ref{eq:14a}) and (\ref{eq:14b}) the bare states are defined in terms of the
adiabatic states for $\tau\rightarrow-\infty$. Then considering the limit for $\tau\rightarrow\infty$, taking into
account (\ref{eq:27}) and the dynamical phase acquired by the third and fourth adiabatic state, we have
\begin{subequations} \label{eq:31}
\begin{align}
&\vert n,e_{1}e_{2}\rangle\rightarrow\vert n,e_{1}e_{2}\rangle, \label{eq:31a} \\ \nonumber \\
&\vert n+1,g_{1}e_{2}\rangle\rightarrow-\vert n+1,e_{1}g_{2}\rangle, \label{eq:31b} \\ \nonumber \\
&\vert n+1,e_{1}g_{2}\rangle\rightarrow \cos(\phi_{n})\vert n+1,g_{1}e_{2}\rangle-\imu\sin(\phi_{n})\vert
n+2,g_{1}g_{2}\rangle \label{eq:31c}, \\ \nonumber \\
&\vert n+2,g_{1}g_{2}\rangle\rightarrow \cos(\phi_{n})\vert n+2,g_{1}g_{2}\rangle-\imu\sin(\phi_{n})\vert
n+1,g_{1}e_{2}\rangle, \label{eq:31d}
\end{align}
\end{subequations}
where the angle $\phi_{n}$ reads
\begin{equation} \label{eq:32}
\phi_{n}=\phi_{4}(\infty)=-\phi_{3}(\infty)=\int_{-\infty}^{\infty}d\tau
E_{4}(\tau).
\end{equation}
These four equations are enough to fully describe the system evolution.

Eq. (\ref{eq:31b}) describes a complete energy transfer between the two atoms. This will
happen without choosing special values for the system parameters provided we ensure the necessary
conditions for adiabatic evolution. This robust energy transfer is a reminiscent of the STIRAP method
\cite{Bergmann1998,Bergmann1995}.
We also note that Eqs. (\ref{eq:31c}) and (\ref{eq:31d}) describe
a conditional entanglement of the second atom to the field mode. If atom 2 is not excited, then it will be entangled to
the field mode. The exact form of the resulting entangled state is defined by $\phi_{n}$ for which we
derive useful asymptotic expressions for various limits.

The first two asymptotic expansions are those for $\delta\ll1$ and $\delta\gg1$. For the latter limit the mixing angle
has a constant value proportional to $\sqrt{n+2}$:
\begin{equation} \label{eq:33}
\phi_{n}\approx4g\sigma\sqrt{(n+2)\pi}\quad\textrm{for}\quad\delta\gg1.
\end{equation}
This asymptotic expansion was derived by using an interpolation for the
integrand in Eq. (\ref{eq:32}). The interpolation function was the sum of two terms,
the first is the limit of $E_{4}(\tau)$ for
$\tau\rightarrow-\infty$ and the second one is the corresponding limit for $\tau\rightarrow\infty$.
For the other limit, $\phi_{n}$ is
\begin{equation} \label{eq:34}
\phi_{n}(\delta)\approx2g\sigma e^{-\delta^{2}}\sqrt{\frac{(6+4n)\pi}{1-\gamma_{n}\delta^{2}}}\quad\textrm{for}\quad
\delta\ll1
\end{equation}
where $\gamma_{n}$ is
\begin{equation} \label{eq:35}
\gamma_{n}=\frac{2((3+2n)^{2}+1)}{(3+2n)^{2}}.
\end{equation}
This asymptotic expansion was derived with the Laplace method
\cite{Wong}. The energy $E_{4}(\tau)$ can be expressed in terms of a function
$w(\tau)$, such that $E_{4}(\tau)=e^{-\ln(w(\tau))}$. For $\delta\ll1$,
the logarithm of $w(\tau)$ can be substituted with a second order
polynomial so that the integral results in Eq. (\ref{eq:34}).

An interesting case of the asymptotic expression for $\phi_{n}$ is the one for $n\gg1$. In this limit one can show that
$\phi_{n}$ has the same dependence with respect to $n$ as the mixing angle in the Jaynes-Cummings model \cite{Scully}:
\begin{equation} \label{eq:36}
\phi_{n}\approx4g\sigma\sqrt{n\pi}\quad\textrm{for}\quad n\gg1.
\end{equation}
This result suggests that for a large number of photons and with adiabatic
evolution, we will have the same kind of dynamics as in the usual
single atom Jaynes-Cummings
model. Furthermore, this Jaynes-Cummings rotation is conditional upon the state of the second atom. This could used for
conditional operations in quantum information or for preparing field states with the use of conditional control. Of course
in this limit the field dynamics will be the same as for the Jaynes-Cummings
model \cite{Scully}.

The mixing angle $\phi_n$ can be tuned with velocity selection methods, so that
  $\sigma\propto1/v$. The sensitivity for these methods 
  is of the order of $\Delta v\sim1\%$  \cite{Varcoe2007}, which corresponds
  to an error of
  the same order for the interaction time. The angle
  $\phi_n$ can also be fine tuned with a Stark shift technique
  previously used in experiments with Rydberg atoms in microcavities
  \cite{Raimond2001,Hagley1997,Osnaghi2001,Rauschenbeutel2000,
    Rauschenbeutel1999,Nogues1999}. 

Finally, before discussing the necessary conditions for using the adiabatic approximation, we comment on the role of
the first atom. The Eqs. (\ref{eq:31a})-(\ref{eq:31d}) can been expressed in terms of a unitary operator $U$. This
matrix can be factorised in terms of two unitary operators. The first one corresponds to a SWAP gate over the atomic
basis with no effect on the field. The subsequent operation is a conditional rotation over the subspace of atom 2 and
the field mode. The condition is that atom 1 must be in its ground state. Thus atom 1 could be seen as the
``control qubit'' during this sequence of unitary operations.
\subsection{Analysis of the adiabatic approximation} \label{sec:3_3}

Throughout sections \ref{sec:3_1} and \ref{sec:3_2} we were assuming that the adiabatic approximation can be used.
For this to be the case, the coupling between the adiabatic states must be smaller than
the corresponding energy splitting \cite{Messiah}
\begin{equation} \label{eq:37}
    Q^{ij}_{n}(\tau,\delta)\equiv\frac{ \left\vert\left\langle\Psi_{i}(\tau)\vert\frac{dH'}{d\tau}\vert\Psi_{j}(\tau)\right\rangle\right\vert}{2\vert
        E'_{i}(\tau)-E'_{j}(\tau)\vert^{2}}\ll g\sigma,
\end{equation}
where $H_{I}'(\tau)=H_{I}(\tau)/g$ and $E'_{j}(\tau)=E_{j}(\tau)/g$.
This choice of parametrisation makes $Q_{n}$ a function of $n$ and
$\delta$ only, since the adiabatic states do not depend on $g$. Thus
in order to quantify the conditions for adiabatic evolution we must
be able to calculate the matrix elements of $Q_{n}$.

Using Eqs. (\ref{eq:13a}), (\ref{eq:13b}) and (\ref{eq:37}), and the fact that the coefficients $A_{\pm}(\tau)$,
$B_{\pm}(\tau)$, $C_{\pm}(\tau)$ and $D_{\pm}(\tau)$ are real, it is easy to show that $Q^{12}_{n}$ and $Q^{34}_{n}$ are zero
for any $n$ and $g$ and arbitrary time $\tau$. For the remaining $Q_{n}^{ij}$ the derivation of analytic expressions is
not simple. For this, we make use of numerical simulations, either by calculating $Q_{n}^{ij}$ or by solving the
Schr\"odinger equation.

\begin{figure}[!t]
  \begin{center}
    \includegraphics{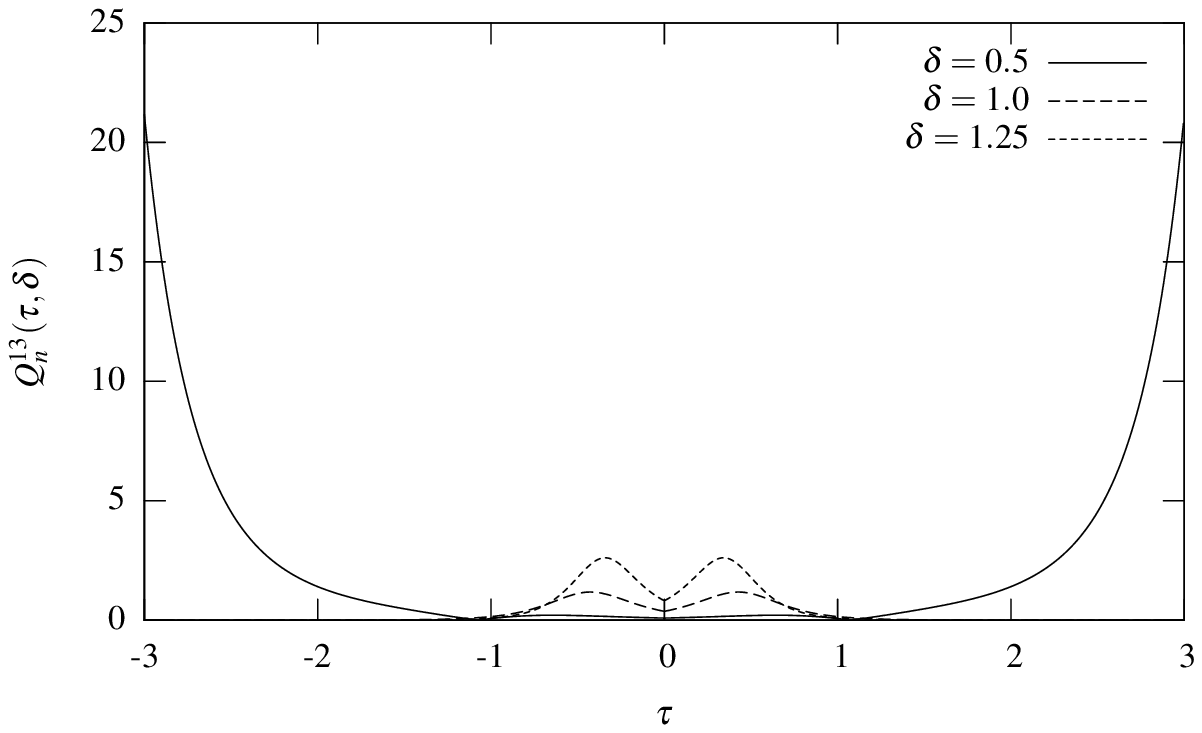}
  \end{center}
  \caption{The matrix element $Q^{13}_{n}(\tau,\delta)$, Eq.
    (\ref{eq:37}), with respect to $\tau$ and $\delta=0.5$, $1.0$ and $1.25$.
    The photon number was $n=0$ and the step $d\tau=0.01$.} \label{fig:3}
\end{figure}

The variation of $Q_{n}$ with respect to $\delta$ and $n$, suggests
that the adiabatic approximation can be used as long as
$\delta\sim1$ and $n$ is kept small. For $n<10$ and $\delta\ll1$, we
have that $Q_{n}\gg1$, figure \ref{fig:3}, whereas for
$\delta\sim1$, $Q_{n}$ is substantially suppressed. Furthermore, we
must ensure that $g\sigma\gg Q_{n}$ at all times, i.e.
\begin{equation} \label{eq:38}
g\sigma\gg\textrm{max}\{Q_{n}(\tau,\delta)\}.
\end{equation}
For $n<10$, the maximum of $Q_{n}$ is of order $1$ and Eq. (\ref{eq:38}) could be
satisfied with $g\sigma\geq10$. If more excitations
are to be added, then the lower bound is shifted
towards higher values. Thus $g\sigma$ must be increased in order to
compensate for the increase in $Q_{n}$.
\begin{figure}[!t]
  \begin{center}
    \includegraphics{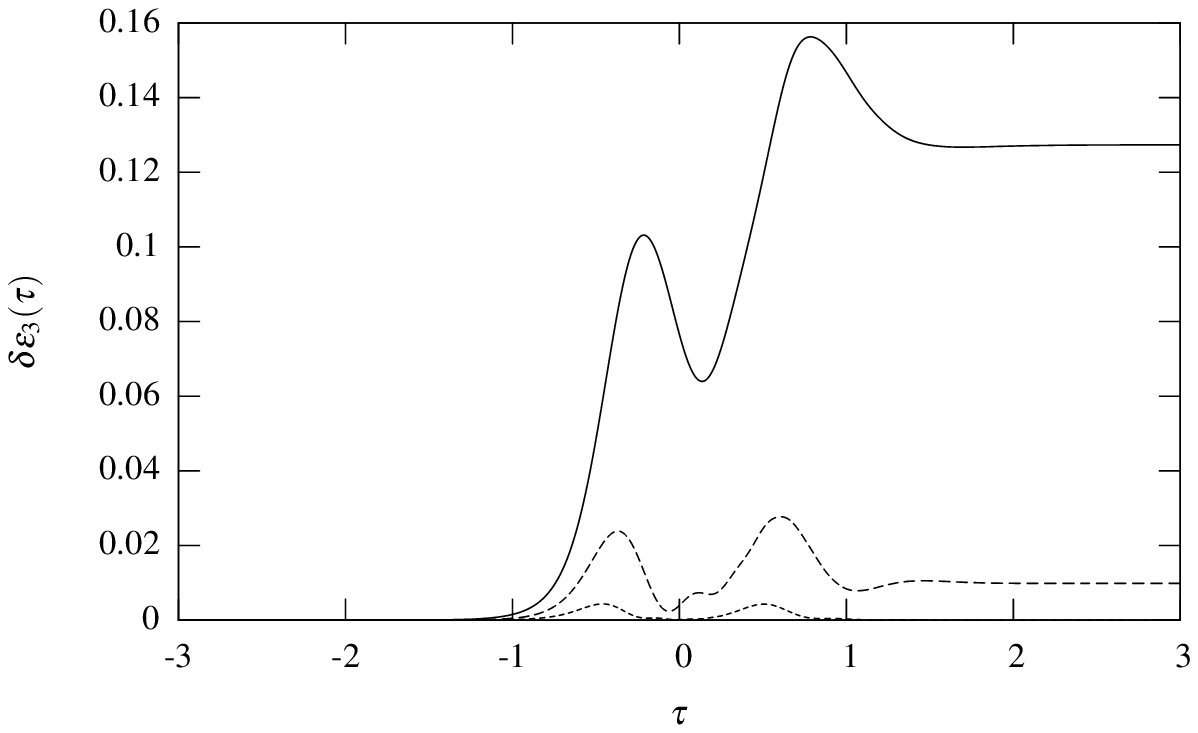}
  \end{center}
  \caption{ The function $\delta\epsilon_{3}(\tau)$ with respect to $\tau$ and different $g\sigma$: $g\sigma=5.0$ (solid), $g\sigma=10.0$ (long dashed) and $g\sigma=20$ (dashed).
    The other parameters are $n=0$ and $\delta=1.0$.}
\label{fig:4}
\end{figure}

These conditions are confirmed by the results obtained after solving
the Schr\"odinger equation. Using as input the adiabatic states, we
integrate the Schr\"odinger equation for a time interval
$\vert\tau\vert\leq\tau_{\scriptscriptstyle 0}$. Then to quantify
the non-adiabatic effects, and understand how they can be suppressed
by proper tuning of the physical parameters, we use the
function
\begin{equation} \label{eq:39}
\delta\epsilon_{j}(\tau)=\left\vert1-\left\vert\left\langle\Psi_{j}(\tau)\vert\Psi(\tau)\right\rangle\right\vert^{2}
\right\vert.
\end{equation}

\begin{figure}[!t]
  \begin{center}
    \includegraphics{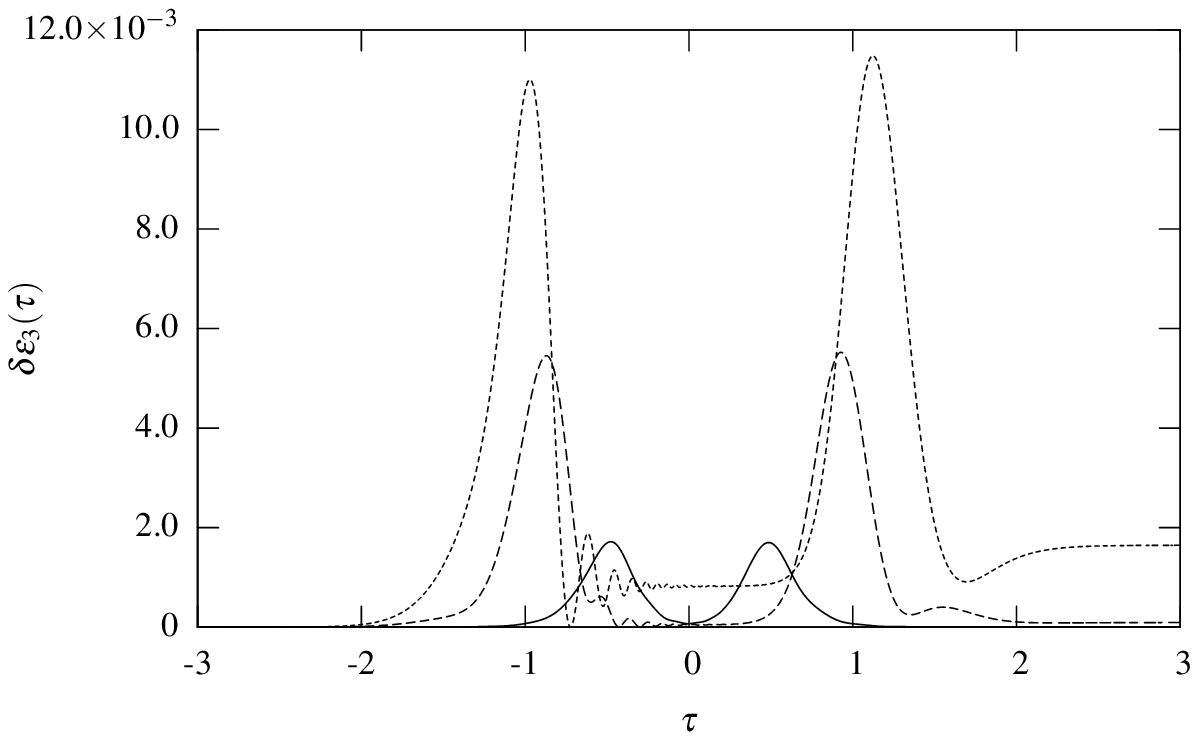}
  \end{center}
  \caption{The function $\delta\epsilon_{3}(\tau)$ with respect to $\tau$ and different $n$: $n=0$ (solid), $n=5$ (long dashed) and $n=10$ (dashed).
    The parameters are $g\sigma=30$ and $\delta=1.0$.
    }
    \label{fig:5}
\end{figure}

What we find from these simulations is that the delay parameter $\delta$ must be of the order of unity, i.e.\
$1\leq\delta\leq1.25$. Furthermore the coupling amplitude $g$ and the interaction time $\sigma$,
must be large enough so that their product exceeds a lower bound which is a function
of the photon number $n$. For $n=0$, this lower bound is approximately equal to $10$, figure \ref{fig:4}, and $g\sigma\gg10$.
For larger $n$, the bound increases, figure \ref{fig:5}, and the coupling amplitude and the interaction time must become
larger in order for the bound to be satisfied.

\begin{figure}[!t]
  \begin{center}
    \includegraphics{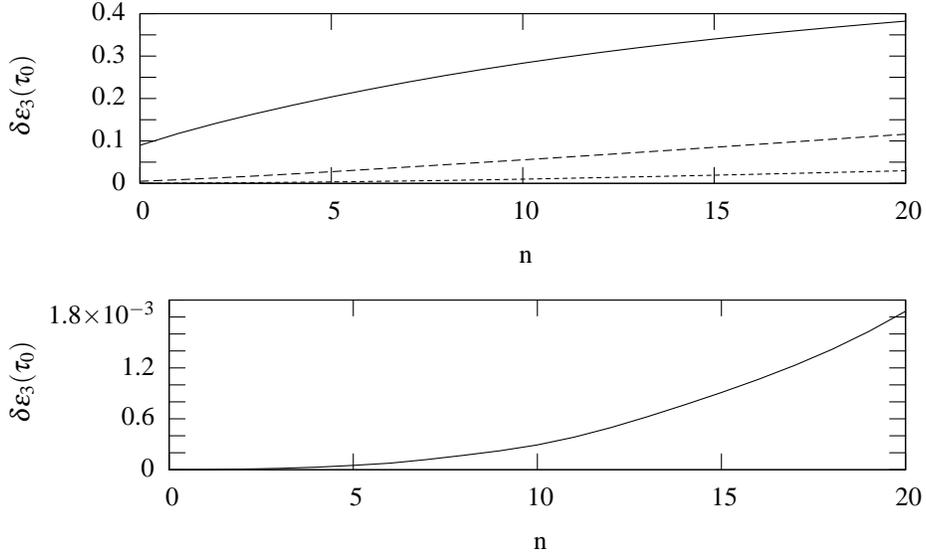}
  \end{center}
  \caption{The function $\delta\epsilon_{3}(\tau)$ for $\tau\rightarrow\infty$ with
    respect to $n$. Top: $g\sigma=10$ (solid), $g\sigma=20$ (long dashed) and
    $g\sigma=30$ (dashed). Bottom: $g\sigma=50$. For both plots we have $\delta=1.2$.
    }
    \label{fig:7}
\end{figure}

The violation of the adiabatic approximation with increasing photon
number is also shown in figure \ref{fig:7}. For increasing values of $n$,
and assuming that the product $g\sigma$ is constant, we see that the adiabatic
approximation will fail. Thus this approximation has a dependence with
respect to $n$. Then if we want the system to evolve adiabatically for larger
photon numbers, a stronger coupling must be chosen.

Physically the requirement $\delta\sim1$, means that while the atoms must both be in
the cavity at some point, they must enter and
exit the cavity at different times, figure \ref{fig:1}. This
demands control over the delay time and the velocities of the atoms.
Furthermore, condition (\ref{eq:38}) requires strong couplings
$g$, and suppression of decoherence to ensure longer interaction times $\sigma$.

A typical optical cavity \cite{Boca2004} has a lifetime of the order of $40$ns while the coupling strength
is approximately $200$MHz. This means that the dimensionless product $g\sigma$ cannot exceed $10$. For a micromaser
cavity \cite{Brune1996}, with quality factor of the order of $10^{8}$ the photon lifetime is approximately $160\mu s$,
with the effective interaction time being equal to $20\mu s$. The coupling is about $150$kHz which gives
$g\sigma\sim3$. For even higher quality factors, e.g. $10^{10}$, the photon lifetime is of the order of $0.1$s which
enables interaction times of the order of $\sigma\sim100\mu s$ \cite{Varcoe2004}. Then with the same coupling, $g\sim150$kHz, we have that
$g\sigma\sim15$ which could be enough for observing adiabatic evolution with
$n\sim1$. This results in sufficiently fast atoms to overcome the barrier
due to the coupling, e.g. for Rb atoms $v\sim100m/s$ \cite{Varcoe2004}. Thus a high quality micromaser
cavity appears to be a good candidate for realising the adiabatic evolution for low photon numbers. Moving
towards higher $n$ would require either higher quality factors or stronger fields.

\section{Applications} \label{sec:4}
\subsection{Atomic entanglement} \label{sec:4_1}
We have already seen in section \ref{sec:3_2} that a single passage of the
atoms through the cavity can entangle the atoms to the cavity. For atomic entanglement,
we would like the cavity to be disentangled. In order to do this
we must prepare the atoms in superpositions of
the form $\alpha_{j}\vert
g_{j}\rangle+\beta_{j}e^{\dot{\imath}\theta_{j}}\vert e_{j}\rangle$ while the
cavity is kept empty
\begin{equation} \label{eq:40}
  \left\vert\Psi_{\textrm{in}}\right\rangle=\vert0\rangle\otimes\left(\alpha_{1}\vert
  g_{1}\rangle+\beta_{1}e^{\dot{\imath}\theta_{1}}\vert e_{1}\rangle\right)
  \otimes\left(\alpha_{2}\vert
  g_{2}\rangle+\beta_{2}e^{\dot{\imath}\theta_{2}}\vert e_{2}\rangle\right).
\end{equation}
The coefficients $\alpha_{j}$ and $\beta_{j}$ are real and $\alpha_{j}^{2}+\beta_{j}^{2}=1$.
After the passage of the atoms through the cavity and assuming, for now, that
$\phi_{-1}$ is arbitrary, the output state will be
\begin{equation} \label{eq:41}
  \left\vert\Psi_{\textrm{out}}\right\rangle=\vert0\rangle\otimes\left\vert\Psi_{en}\right\rangle-
  \imu\beta_{1}
  \alpha_{2}\sin(\phi_{-1})e^{\imu\theta_{1}}\vert1\rangle\otimes\vert
  g_{1}g_{2}\rangle,
\end{equation}
where the atomic state $\left\vert\Psi_{\textrm{en}}\right\rangle$ reads
\begin{eqnarray}
\nonumber
  \left\vert\Psi_{\textrm{en}}\right\rangle&=&\alpha_{1}\vert g_{2}\rangle\left(\alpha_{2}\vert
  g_{1}\rangle-\beta_{2}e^{\imu\theta_{2}}\vert e_{1}\rangle\right)\\ & &+\beta_{1}e^{\imu
    \theta_{1}}\vert e_{2}\rangle\left(\alpha_{2}\cos(\phi_{-1})\vert g_{1}\rangle
  +\beta_{2}e^{\imu\theta_{2}}\vert e_{1}\rangle\right). \label{eq:42}
\end{eqnarray}

Thus, if the mixing angle $\phi_{-1}$ is not equal to $(2m+1)\pi$, where
$m=0,1,2\dots$, the atomic state $\left\vert\Psi_{\textrm{en}}\right\rangle$
will be an entangled state of the two atoms. The probability $P_{\textrm{en}}$ for this
state is
\begin{equation} \label{eq:43}
  P_{\textrm{en}}=1-\beta_{1}^{2}\alpha_{2}^{2}\sin^{2}(\phi_{-1}).
\end{equation}
If we want to generate a maximally entangled state, then we must ensure that
$P_{\textrm{en}}=1$ by choosing $\phi_{-1}=2\pi$. Furthermore $\alpha_{j}$ and
$\beta_{j}$ must be  $1/\sqrt{2}$ and the phase angle $\theta_{2}$ is
zero. Then the resulting state is
\begin{equation} \label{eq:44}
  \left\vert\Psi_{\textrm{en}}\right\rangle=\frac{1}{2}\left(\vert
  g_{2}\rangle\left(\vert g_{1}\rangle-\vert e_{1}\rangle\right)+e^{\imu\theta_{1}}\vert
  e_{2}\rangle\left(\vert g_{1}\rangle+\vert e_{1}\rangle\right)\right),
\end{equation}
and a Hadamard rotation $H$ over the subspace of atom 1 \cite{Nielsen} must be applied
\begin{equation} \label{eq:45}
  H\vert g_{1}\rangle=\frac{1}{\sqrt{2}}\big(\vert g_{1}\rangle+\vert
  e_{1}\rangle\big),\qquad H\vert
  e_{1}\rangle=\frac{1}{\sqrt{2}}\big(\vert g_{1}\rangle-\vert
  e_{1}\rangle\big),
\end{equation}
to get the maximally entangled state
\begin{equation} \label{eq:46}
  \left\vert\Psi_{\textrm{en}}\right\rangle=\frac{1}{\sqrt{2}}\left(e^{\imu\theta_{1}}\vert
  g_{1}e_{2}\rangle+ \vert e_{1}g_{2}\rangle\right).
\end{equation}
\subsection{State mapping} \label{sec:4_2}
Assume now that the two atoms are prepared in the following superposition
\begin{equation} \label{eq:47}
  \vert g_{1}\rangle\left(\alpha\vert g_{2}\rangle+\beta\vert
  e_{2}\rangle\right),
\end{equation}
and the cavity is in the vacuum state, then the
state of the system after the passage of the atoms through the cavity will be
\begin{equation} \label{eq:48}
  \vert 0,g_{2}\rangle\left(\alpha\vert g_{1}\rangle-\beta\vert
  e_{1}\rangle\right).
\end{equation}
Applying a single rotation on atom 1, $\vert
e_{1}\rangle\rightarrow-\vert e_{1}\rangle$, the final state of atom 1 reads
\begin{equation} \label{eq:49}
  \alpha\vert g_{1}\rangle+\beta\vert e_{1}\rangle.
\end{equation}
Thus the state of atom 2 can be mapped into atom 1. The whole scheme is simple
and robust since there is no need for exact control over the interaction time,
in contrast to other proposals \cite{Zheng2005}.

In addition to this scheme, one can implement state mapping between the atoms and the
cavity, and vice versa. If the cavity is prepared in the state
$\alpha\vert0\rangle+\beta\vert1\rangle$, the two atoms are in their ground
states, and the mixing angle is $\phi_{-1}=\pi/2$, then the resulting state
for the second atom will be $\alpha\vert g_{2}\rangle-\imu\beta\vert
e_{2}\rangle$ whereas the cavity and the first atom are not excited. If the
 single qubit rotation $\vert e_{2}\rangle\rightarrow\imu\vert e_{2}\rangle$
 is applied, then the resulting state for atom 2 will be
\begin{equation} \label{eq:50}
  \alpha\vert g_{2}\rangle+\beta\vert e_{2}\rangle.
\end{equation}
If now the system is prepared in the state $\vert0,g_{2}\rangle\left(\alpha\vert g_{1}\rangle+\beta\vert
e_{1}\rangle\right)$, and we apply the transformation $\vert
e_{1}\rangle\rightarrow\imu\vert e_{1}\rangle$, and send the atoms through the
cavity with $\phi_{-1}=\pi/2$, then the state of atom 1 will be mapped onto the
cavity mode, i.e.\ the system state becomes
\begin{equation} \label{eq:51}
  \left(\alpha\vert0\rangle+\beta\vert1\rangle\right)\vert g_{1},g_{2}\rangle.
\end{equation}

\subsection{SWAP and C-NOT gates} \label{sec:4_3}

We conclude the analysis of the current section by presenting two setups for implementing
a SWAP and a C-NOT gate. The first can be realised using an empty
cavity and choosing $\phi_{-1}=\pi$. From Eqs.
(\ref{eq:31a})-(\ref{eq:31c}) we have
\begin{eqnarray}
\nonumber \vert e_{1}e_{2}\rangle\rightarrow\vert e_{1}e_{2}\rangle,
&& \vert g_{1}e_{2}\rangle\rightarrow-\vert e_{1}g_{2}\rangle,
\\ \nonumber \vert g_{1}g_{2}\rangle\rightarrow\vert
g_{1}g_{2}\rangle, && \vert e_{1}g_{2}\rangle\rightarrow-\vert
g_{1}e_{2}\rangle,
\end{eqnarray}
with the cavity remaining empty. Thus the model in hand could be
used to implement a SWAP gate, without using any additional
components.

In contrast to this, the realisation of a C-NOT gate requires the use of
additional components. We consider a row of two empty cavities, with
the angle $\phi_{-1}$ being $2\pi$ and $\pi$ respectively. If a pair
of atoms crosses through these two cavities, then the output is
\begin{subequations} \label{eq:52}
\begin{align}
\vert e_{1}e_{2}\rangle\rightarrow\vert e_{1}e_{2}\rangle, &\qquad \vert
g_{1}e_{2}\rangle\rightarrow\vert g_{1}e_{2}\rangle, \label{eq:52a} \\
\vert g_{1}g_{2}\rangle\rightarrow\vert g_{1}g_{2}\rangle, &\qquad \vert
e_{1}g_{2}\rangle\rightarrow-\vert e_{1}g_{2}\rangle. \label{eq:52b}
\end{align}
\end{subequations}
The use of an array of two cavities is allowed since neither cavity is excited
after the atoms have passed through. In addition to this, attention must be
paid, so that both atoms exit the first cavity, before entering the second one.

Eqs. (\ref{eq:52a}) and (\ref{eq:52b}) correspond to a phase
gate. This can be combined with two Hadamard gates
to produce a C-NOT gate. For the current setup, the
Hadamard rotation is performed over the qubit space represented by
atom 1, Eq. (\ref{eq:45}), and is applied before and after the atoms
have crossed the cavities.
Combining Eqs. (\ref{eq:45}), (\ref{eq:52a}) and (\ref{eq:52b})
gives the following input-output table
\begin{eqnarray}
\nonumber&\vert e_{1}e_{2}\rangle\rightarrow\vert e_{1}e_{2}\rangle,
\\ \nonumber&\vert g_{1}e_{2}\rangle\rightarrow\vert
g_{1}e_{2}\rangle, \\ \nonumber&\vert
e_{1}g_{2}\rangle\rightarrow\vert g_{1}g_{2}\rangle, \\ \nonumber
&\vert g_{1}g_{2}\rangle\rightarrow \vert e_{1}g_{2}\rangle,
\end{eqnarray}
which corresponds to a
C-NOT gate, with atom 2 being the
control qubit and atom 1 the target qubit.

\begin{figure}[!t]
  \begin{center}
    \includegraphics{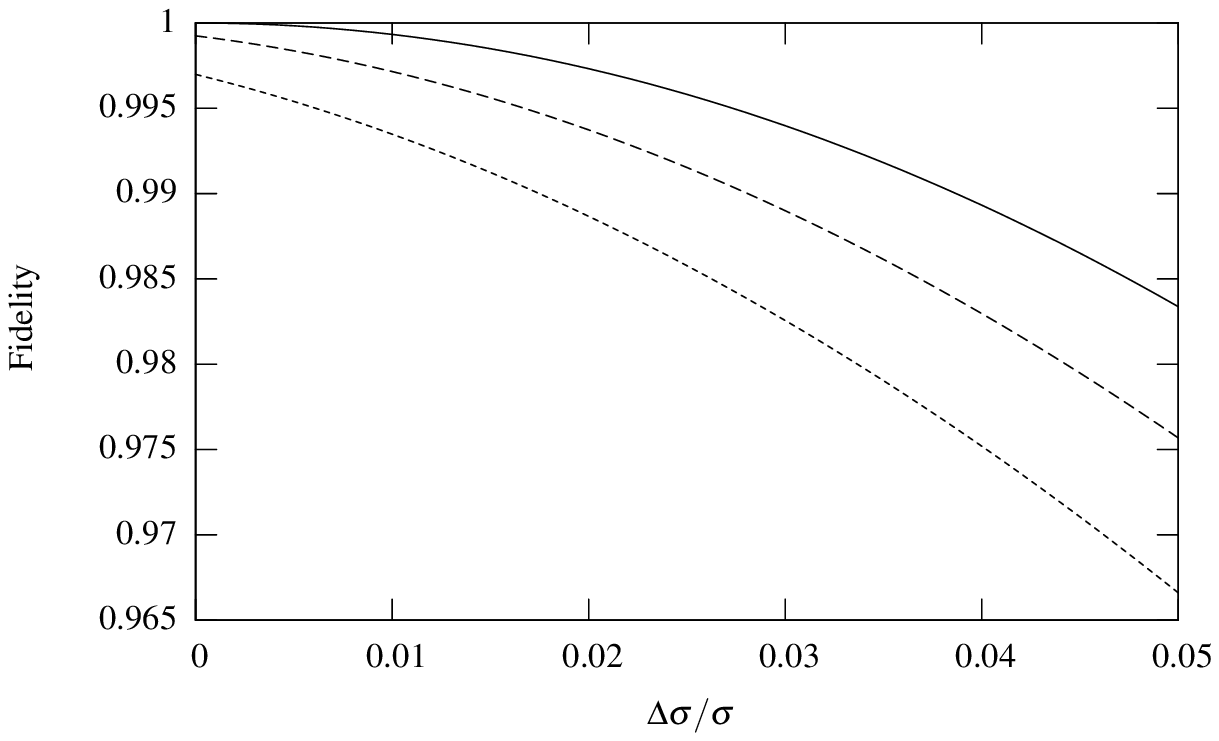}
  \end{center}
  \caption{The fidelity for a maximally entangled state with respect to
    variations in the interaction time $\Delta\sigma$ and different variations in the
  delay time: $0$ (solid), $0.05\Delta t$ (dashed) and $0.1\Delta t$
  (dot). The parameters
  $\sigma$ and $\Delta t$ correspond to $\delta=1.0$, $\phi_{-1}=2m\pi$ and $\theta_{1}=0$. } \label{fig:6}
\end{figure}

With the exception of the state mapping between atoms, which is a robust
scheme, the other proposed applications in this section could be characterised as \emph{fairly}
robust. The fidelities for these applications can exceed $99\%$ with proper
control of the physical parameters such as interaction time $\sigma$ and delay time $\Delta t$.
For example, in figure \ref{fig:6} the fidelity for a
maximally entangled state is plotted with respect to $\Delta\sigma$ for various
errors in the delay time. From this we see that the fidelity can
be greater than $97\%$ if the error in $\sigma$ is below $5\%$ while the
corresponding error for the delay time can be as high as $10\%$. With better
tuning of the interaction time and after suppressing the error $\Delta\sigma$
below $2\%$ the fidelity can exceed $99\%$.

In general for all the given applications the corresponding fidelity is less sensitive
to errors in the delay time in contrast to errors in the interaction time
$\sigma$. This means that the tuning of the interaction time has to be more
accurate than the corresponding tuning for the delay time. Furthermore, for
some of the applications, such as  mapping the cavity state
$(\vert0\rangle+\vert1\rangle)/\sqrt{2}$ onto atom 2, the threshold for the
error in $\sigma$ could be as high as $10\%$ with a resulting fidelity equal
to $99\%$.

Decoherence effects related to spontaneous emission from the atoms and
the dissipation of photons
from the cavity, are detrimental to the proposed applications. The
adiabatic condition, Eq. (\ref{eq:38}), suggests that the mean
interaction time can be substantially decreased while increasing the coupling strength $g$
to ensure adiabatic evolution. This will allow fast implementation of the
proposed applications, reducing the probability for potential failures due to
decoherence.

Finally, we note that the  implementation of the single atom rotations
can be carried out with classical EM
fields \cite{Yamaguchi2002}. For the rotations to be successful, we must ensure
that the spatial displacement between the atoms is sufficiently large. This
will prevent interactions between the fields and both atoms when only one atom
must couple to the EM fields.

\section{Conclusions} \label{sec:Conl}

In this paper, we consider the situation where two atoms interact with a single mode cavity field via sequential time
dependent couplings. We focused on the adiabatic limit, but were also able to show that an energy crossing takes place in the
vicinity of a temporal degeneracy. This takes place because the degenerate adiabatic states do not couple to each other
near the degeneracy point. This effect, together with the symmetry properties of the adiabatic spectrum, makes the system
fairly robust. The only parameter which must be controlled is a mixing angle
$\phi_{n}$. This angle defines the degree of conditional entanglement
between the atoms and the cavity. In the limit of large photon number,
the mixing angle has the same dependence with respect to the photon number as
in the usual single atom Jaynes-Cummings model. Thus in this limit we could say that the system
behaves like a conditional Jaynes-Cummings system.

A significant difference from previously proposed theoretical schemes and
  experiments carried out
  \cite{Zheng2005,Zheng2000,Jane2002,You2003b,Plenio1999,Beige2000a,Beige2000b,Duan2003,Duan2004,Duan2005,Duan2003A,Raimond2001,Hagley1997,Osnaghi2001,Rauschenbeutel2000,Rauschenbeutel1999,Nogues1999},
  is the energy crossing which is seen in figure \ref{fig:1}. This feature enables the robust disentanglement of the atoms from the cavity, after the evolution of the system is
  completed.
Based on this, we have proposed methods for entangling atoms and mapping quantum states
between atoms and cavities and experimental setups for implementing a SWAP and
a C-NOT gate. The proposed schemes are fairly robust, with fidelities up to
$99\%$. This robustness is especially evidenced in the delay time since errors
of up to $10\%$
have minor effect on the fidelities of the proposed applications. Furthermore,
the fact that the interaction time can be substantially
decreased makes these applications less sensitive to decoherence effects.

What makes this system fairly robust is the symmetric structure of the adiabatic spectrum and the energy crossing.
Both of these features are sensitive to variations of the detuning, and to changes of the ratio of the
coupling amplitudes. If the coupling amplitudes have a ratio that is slightly different from unity, then the system
evolution will be significantly different. This should be expected since a second mixing angle appears in Eqs.
(\ref{eq:31}), and one must be able to control the ratio $g_{1}/g_{2}$ in order to have the desired output. Furthermore,
the existence of a detuning between the atoms and the field will lift the temporal degeneracy, preventing the energy
crossing from occurring. The detuning also affects the structure of the spectrum and its symmetry properties.

The system is vulnerable to
decoherence effects. Since we are interested in the adiabatic limit, the
interaction time is expected to be large. From our analysis we conclude
that the adiabatic condition allows shorter interaction times by simply
increasing the coupling strength. In this way a faster process can be realised,
reducing potential failures due to decoherence. Based on current experiments,
either with optical or micromaser cavities, we find that a high quality
micromaser cavity could be used for realising the proposed applications.

\acknowledgments{BMG acknowledges support from the Leverhulme Trust.}

\appendix
\section{Adiabatic states}
The adiabatic states are derived after solving Eq. (\ref{eq:5}). The normalised adiabatic states are
given in (\ref{eq:13a}) and (\ref{eq:13b}) where the coefficients $A_{\pm}(\tau)$, $B_{\pm}(\tau)$, $C_{\pm}(\tau)$
and $D_{\pm}(\tau)$ are
\begin{subequations} \label{eq:53}
\begin{align}
A_{\pm}(\tau)&=\pm\sqrt{\frac{4(n+1)(n+2)\eta_{\scriptscriptstyle
1}^{2}\eta_{\scriptscriptstyle
2}^{2}}{F_{n}^{2}(\tau)\pm(\eta_{\scriptscriptstyle
1}^{2}+\eta_{\scriptscriptstyle 2}^{2})F_{n}(\tau)}}, 
\label{eq:53a} \\ \nonumber \\
B_{\pm}(\tau)&=\frac{D_{\pm}(\tau)E_{\pm}(\tau)\left(\eta_{\scriptscriptstyle
1}^{2}-\eta_{\scriptscriptstyle 2}^{2}\mp
F_{n}(\tau)\right)}{2\eta_{\scriptscriptstyle
2}\sqrt{n+2}\left(E_{\pm}^{2}(\tau)+(n+1)(\eta_{\scriptscriptstyle
1}^{2}-\eta_{\scriptscriptstyle 2}^{2})\right)},
\label{eq:53b} \\ \nonumber \\
C_{\pm}(\tau)&=\frac{D_{\pm}(\tau)E_{\pm}(\tau)\eta_{\scriptscriptstyle
1}(3+2n)}{\sqrt{n+2}\left(E_{\pm}^{2}(\tau)+(n+1)(\eta_{\scriptscriptstyle
1}^{2}-\eta_{\scriptscriptstyle 2}^{2})\right)}, \label{eq:53c} \\ \nonumber \\
D_{\pm}(\tau)&=\frac{1}{2}\sqrt{1\pm\frac{\eta_{\scriptscriptstyle
1}^{2}+\eta_{\scriptscriptstyle 2}^{2}}{F_{n}(\tau)}}. \label{eq:53d}
\end{align}
\end{subequations}
Similar expressions were previously derived in the context of pair effects in single atom micromasers \cite{Mahmood1987}.

\bibliography{Paper.bbl}
\end{document}